\documentclass[aps,pre,showpacs%
        ,amsmath%
        ,superscriptaddress%
        ,reprint%
        ,nofootinbib
        ,preprintnumbers
        ]{revtex4-1}

\usepackage[pdftex]{graphicx} 

\usepackage{amssymb}

\def\ack{\section*{Acknowledgements}}

\newcommand{\ket}[1]{|{}#1{}\rangle}
\newcommand{\bra}[1]{\langle{}#1{}|}

\begin{document}

\title{Complete population inversion of Bose particles by an adiabatic cycle}

%

\author{Atushi Tanaka}
\homepage[]{\tt http://researchmap.jp/tanaka-atushi/}
\affiliation{Department of Physics, Tokyo Metropolitan University, Hachioji, Tokyo 192-0397, Japan}

\author{Taksu Cheon}
\homepage[]{\tt http://researchmap.jp/T_Zen/}
\affiliation{Laboratory of Physics, Kochi University of Technology, Tosa Yamada, Kochi 782-8502, Japan}

\begin{abstract}
We show that an adiabatic cycle excites Bose particles confined in a one-dimensional box. During the adiabatic cycle, a wall described by a $\delta$-shaped potential is applied and its strength and position are slowly varied. When the system is initially prepared in the ground state, namely, in the zero-temperature equilibrium state, the adiabatic cycle brings all bosons into the first excited one-particle state, leaving the system in a nonequilibrium state.
The absorbed energy during the cycle 
is proportional to the number of bosons.
\end{abstract}


\maketitle 

\section{Introduction}

The population inversion of quantum states has been investigated for its application to lasing~\cite{IntroQO}. 
The quantum control of atoms and molecules also has been investigated to realize the population inversion~\cite{ShoreBook,ShapiroBrumerBook}.
Recently, studies of the super-Tonks-Girardeau gas, which also involves
the population inversion, has 
attracted a lot of attention in both experimental and theoretical studies of 
nonequilibrium cold 
atoms~\cite{Astrakharchik-PRL-95-19407,Olshanii-PRL-81-938,Haller-Science-325-1224,Guan-IJMPB-28-1430015}.
In the super-Tonks-Girardeau gas, which may be described by the Lieb-Liniger model~\cite{Lieb-PR-130-15} with strongly attractive interaction, the population inversion is created through an ``adiabatic'' process, where the interaction
strength is suddenly flipped from infinitely repulsive
to infinitely attractive~\cite{Haller-Science-325-1224, Girardeau-JMP-1-516}.

Such a population inversion can be induced even by an adiabatic cycle, which 
can be obtained with an extension of the adiabatic process that connects 
Tonks-Girardeau and super-Tonks-Girardeau gases both to weaker repulsive and 
weaker attractive regime. The repetitions of this adiabatic cycle
transform
the ground state of non-interacting bosons 
into
their higher excited 
states 
and achieve
the population inversion~\cite{Yonezawa-PRA-87-062113}.
%
This is counterintuitive, since there is no external field to 
drive the final state of the bosons away from the initial state.

There 
has been studies of the excitation of quantum systems
by adiabatic cycles,
which is referred to as 
the exotic quantum 
holonomy~\cite{Cheon-PLA-248-285,Tanaka-PRL-98-160407,Tanaka-PLA-379-1693}.
We also mention, in studies of atomic and molecular systems under the oscillating field, that an adiabatic cycle involving a level crossing may excite a quantum system~\cite{Vitanov-ARPC-52-763,Guerin-PRA-63-031403}.

In this paper, we examine an adiabatic cycle that excites a system 
consisting of Bose particles confined in a one-dimensional box.
During the cycle, we vary an additional wall adiabatically, while
the interparticle interaction is kept fixed. This is in contrast 
to the scheme described in 
Refs.~\cite{Haller-Science-325-1224,Yonezawa-PRA-87-062113}, where
the interaction strength between Bose particles is an effective
adiabatic parameter.
In this study, we suppose that the wall is described by 
a $\delta$-function shaped 
potential~\cite{Flugge1971,Ushveridze-JPA-21-955,Kasumie-arXiv-151007854}.
We show
that the first excited one-particle state is occupied by 
all the
bosons to achieve the population inversion completely, if the system is prepared to be in the ground state.
Namely, the energy gained by the bosons during the adiabatic cycle is proportional to the number of 
bosons.

\section{A particle in a box
  with a $\delta$-wall}

In order to examine $N$ Bose particles in a one-dimensional box
with an additional $\delta$-wall,
we review the single particle case, i.e., $N=1$~\cite{Kasumie-arXiv-151007854},
where the system is described by the Hamiltonian
\begin{equation}
  \label{eq:def_H1}
  H(g,X) = \frac{p^2}{2m} + V(x) + g\delta(x-X)
  ,
\end{equation}
where $m$ is 
the particle mass,
$V(x)$ 
is the confinement potential, and $g$ and $X$ are
the strength and position of $\delta$-wall.
In particular, we assume that $V(x)$ describes an infinite square well
with the length $L$, i.e., 
$V(x)=0$ for $0 < x < L$
and $V(x)=\infty$ otherwise~\cite{Flugge1971,Ushveridze-JPA-21-955}.
%

%
\begin{figure}[b]
  \centerline{%
    \includegraphics[%
    	width=0.45\textwidth
        ]{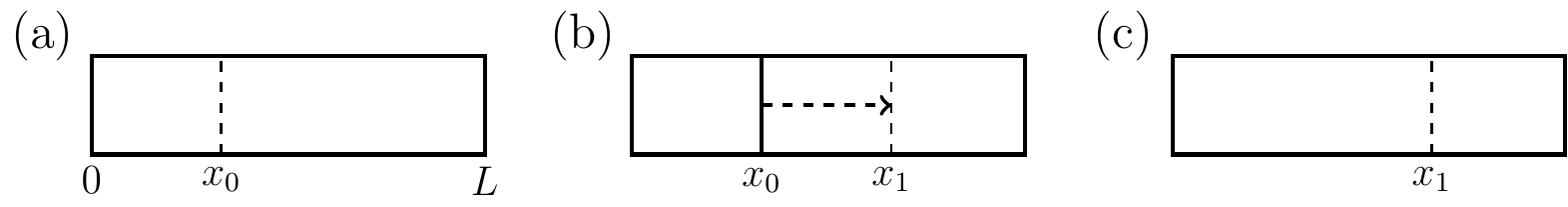}}
  \caption{%
    The adiabatic cycle $C$
    of a one-dimensional box, which contains 
    Bose particles. The strength and the position of an additional 
    $\delta$-wall is adiabatically varied during $C$.
    The cycle consists of three processes
    $C_I$, $C_{II}$ and $C_{III}$.
    (a) In the first process $C_{I}$,
    the $\delta$-wall is placed at $x_0$
    and its strength $g$ is adiabatically increased
    from $0$ to $\infty$.
    (b) In the second process $C_{II}$,
    the position of the impenetrable wall is adiabatically
    moved from $x_0$ to $x_1$, while keeping its strength $\infty$.
    (c) The final process $C_{III}$, 
    the $\delta$-wall at $x_1$ is 
    adiabatically turned off.
  }
\label{fig:box}
\end{figure}
We introduce an adiabatic cycle $C$, which consists of three adiabatic
processes
$C_{I}$, $C_{II}$ and $C_{III}$, as shown 
in Figure~\ref{fig:box}.
We suppose that the $\delta$-wall is initially absent, i.e., $g=0$
in \eqref{eq:def_H1}, and that the system is in a stationary state 
initially.
In the first part of $C$, which will be called as $C_{I}$,
an impenetrable wall is inserted at $x_0$ adiabatically. 
In terms of the $\delta$-wall, the strength $g$ is
slowly increased from $0$ to $\infty$, while its position $X$ is
fixed at $x_0$ during $C_{I}$.
Subsequently, in the second part $C_{II}$,
the position $X$ of the impenetrable wall is adiabatically changed from $x_0$
to $x_1$. In the last part $C_{III}$, the $\delta$-wall at $X=x_1$ 
is adiabatically turned off. At the end of the cycle $C$, the $\delta$-wall
has no effect, again.
%

\begin{figure}
  \centerline{\includegraphics[%
    	height=0.20\textheight
        ]{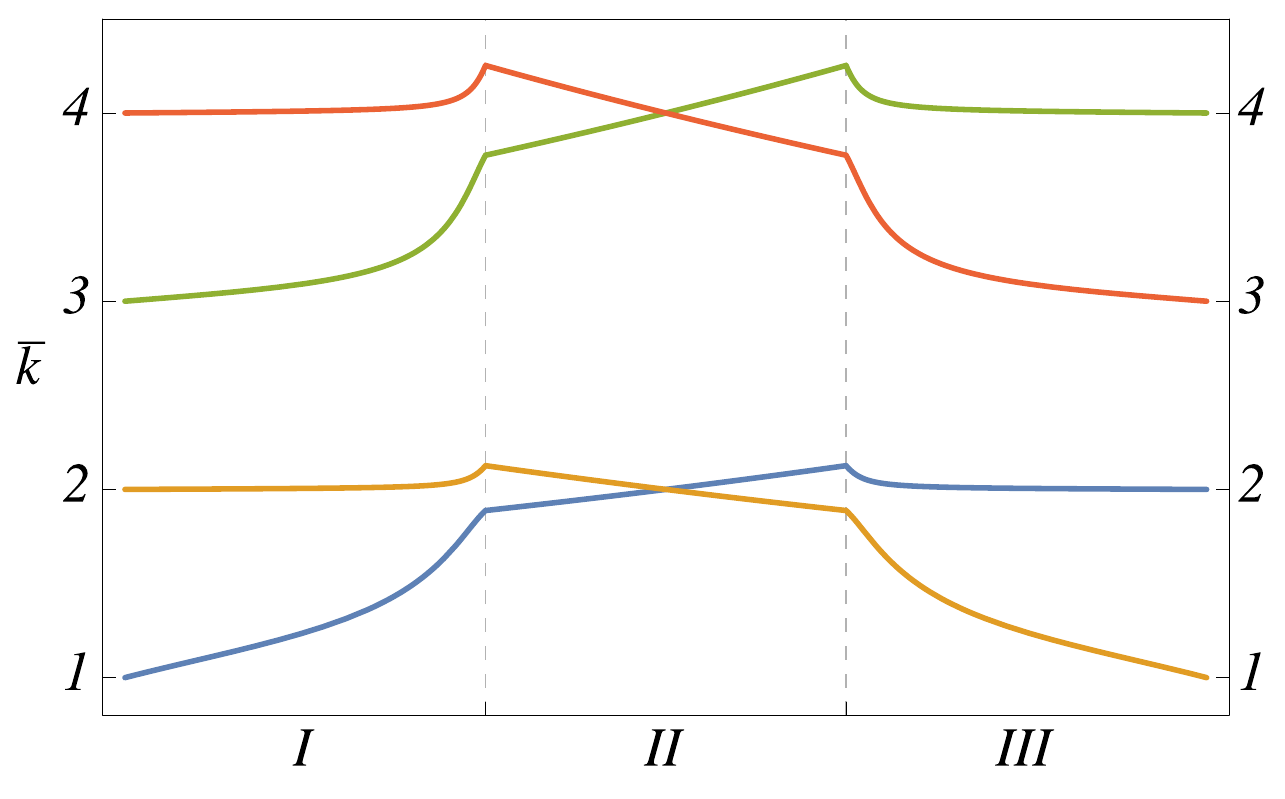}}
  \caption{%
    Parametric evolution of eigenenergies with $N=1$
    along the cycle $C$, which consists of $C_I$ (left part), 
    $C_{II}$ (middle part) and $C_{III}$ (right part).
    The eigenenergies are depicted by 
    their normalized wavenumber $\bar{k}$ \eqref{eq:def_bar_k}.
    The initial states corresponding to these four levels 
    are $\ket{n}$, whose quantum number $n$
    coincides with $\bar{k}$ at the initial point of the cycle.
    We note that these levels correspond to 
    the $N$-particle states 
    $\ket{n^{\otimes N}}$ \eqref{eq:n_otimesN}
    of the noninteracting Bose particles ($n=1,\dots{},4$).
    We set $x_0 = 0.4703L$ and $x_1 = L - x_0$.
  }
  \label{fig:N1}
\end{figure}

In Figure~\ref{fig:N1}, 
we depict the parametric dependence of 
eigenenergies of
the single-particle Hamiltonian $H(g,X)$ \eqref{eq:def_H1}
along $C$.
Throughout this manuscript, we indicate the eigenenergy $E$
using a normalized wavenumber $\bar{k}$
\begin{equation}
  \label{eq:def_bar_k}
  \bar{k}\equiv\sqrt{\frac{E}{N\epsilon}}
  ,
\end{equation}
where $\epsilon = (\hbar\pi/L)^2/(2m)$ is
the ground eigenenergy of the particle in the infinite square well.

The adiabatic time evolution of the single-particle system along
$C$ depends on $x_0$ and $x_1$.
In the following, we explain the case 
$\frac{2}{5}L < x_0 < \frac{1}{2}L < x_1 < \frac{3}{5}L$,
which may be explained from Figure~\ref{fig:N1}.
A more rigorous argument is found in Ref.~\cite{Kasumie-arXiv-151007854}.

First, let us consider the case that the initial state is
the ground state $\ket{1(g=0,X=x_0)}$
of the particle in the infinite square well,
where $\ket{n(g,X)}$ denotes the $n$-th adiabatic eigenstate 
of $H(g,X)$ during processes $C_I$ and 
$C_{III}$.
We will omit to indicate $(g,X)$ in the following.
After the completion of $C_I$, the state vector
arrives at $\ket{R_1}$, the ground state of the right well,
since we choose the right well in $C_{II}$ is slightly larger
than the left well.
During $C_{II}$, there occurs a spectral degeneracy
between $\ket{R_1}$ and $\ket{L_1}$, the ground state of the 
left well. This is because the size of the left (right) well is
increasing (decreasing) during $C_{II}$, and these sizes 
coincide at $X = L/2$. At the end of $C_{II}$, $\ket{R_1}$
becomes the first excited state, which adiabatically continued
to $\ket{2}$, which is the second excited state
of the particle in the infinite square well,
through $C_{III}$. 
Hence the ``population inversion'' in the single-particle system
occurs if the system is prepared to be in the ground state initially.

Second, we examine the case that the initial state is the first
excited state $\ket{2}$,
which offers the ``inverse'' of the population inversion.
Through the adiabatic cycle $C$, the system arrives 
at $\ket{L_1}$ after the completion of $C_1$, and then
arrives at $\ket{1}$ at the end of the cycle $C$.
Namely, either $\ket{1}$ and $\ket{2}$
return to the initial states
after the completion of the 
adiabatic cycle $C$ twice.

Third, let us examine the cases that 
the initial states are $\ket{3}$ and $\ket{4}$,
which are the first and second excited state, respectively.
Now $C$ induces an interchange of these two states, through
the intermediate states $\ket{R_2}$ and $\ket{L_2}$, which are
localized the right and left well during the process II.

A similar interchange of initial eigenstates occurs as a result
of the adiabatic cycle $C$, as long as we choose 
$x_0$ and $x_1$ appropriately. 
In general, the level crossing of the one-particle 
Hamiltonian~\eqref{eq:def_H1} during the process $C_{II}$ 
plays an important role to determine which pairs of eigenstates
are interchanged by $C$, while there is no level crossing
generically during the processes $
C_{I}$ and $C_{III}$~\cite{Kasumie-arXiv-151007854}.

We make a remark on the stability of the present scheme for the one-body 
population inversion. A crucial point is the stability of
the adiabatic time evolution across the level crossing during $C_{II}$.
The level crossing may be lifted due to an imperfection of
the impenetrable wall, i.e., the $\delta$-wall with an infinite strength.
If the level splitting is small enough, we may employ the diabatic
process around the avoided crossing to realize the
one-body population inversion. 
It has been shown
that an open diabatic process made of time-dependent potential well
produce the second excited state from the ground state of a bose 
particle~\cite{Karkuszewski-PRA-63-061601}.
This diabatic scheme is applied to create collective excitations
of interacting bosons~\cite{Karkuszewski-PRA-63-061601,Damski-PRA-65-013604,Karkuszewski-EPJD-21-251}.

\section{Non-interacting Bosons}
We examine the case that the number of
the Bose particles is $N$, assuming the absence of interparticle interaction.
It is straightforward to extend the above result for $N=1$, once 
we restrict the case that $N$ bosons initially occupies 
the one-particle state $\ket{n}$. Hence the system is in 
an adiabatic state of the $N$ 
bosons
\begin{equation}
  \label{eq:n_otimesN}
  \ket{n^{\otimes N}}\equiv \ket{nn{\ldots}n}
  ,
\end{equation}
where the one-particle adiabatic state $\ket{n}$ is occupied by $N$ bosons,
during $C_{I}$ and $C_{III}$. 

If there is no interparticle interactions,
the parametric evolution of averaged wavenumber $\bar{k}$ 
\eqref{eq:def_bar_k} for the adiabatic $N$-particle
state agree with the one of the single-particle system.
This suggests that the adiabatic cycle $C$ of the $N$-particle
system with no interaction
delivers 
the ground state $\ket{1^{\otimes N}}$
to the excited state $\ket{2^{\otimes N}}$,
i.e., the complete population inversion,
as is seen in Figure~\ref{fig:N1}.
The energy that the particles acquire during the cycle $C$
is proportional to the number of the particles.

\section{Interacting Bosons}
We examine the adiabatic cycle $C$ for 
$N$
interacting Bose
particles. We mainly examine the case that the system is initially
in the ground state. 
In order to confirm that the $N$-particle population inversion
really occurs, we need to examine the effect of the interparticle 
interaction. 

We 
assume that the interparticle interaction  ${V}$ consists of
two-body contact interactions. 
Namely, 
we suppose that ${V}$ takes the
following form  
\begin{equation}
  V(x_1, x_2, \ldots, x_N) = \lambda\sum_{\langle{}i,j\rangle}\delta(x_i - x_j),
\end{equation}
where $\lambda$ is the interaction strength, and
the summation is taken over pairs.

We 
also
assume that the interparticle interaction is
weak enough so that the topology of the parametric dependence of 
eigenenergy remain unchanged, except 
around
the level 
crossing points
of the noninteracting 
bosons.
Namely, when the gaps of the eigenenergies between neighboring levels in
the noninteracting system are 
larger than a constant value, 
the interparticle interaction
shifts the eigenenergy at most ${\cal O}(\lambda)$, according
to the standard perturbation theory.
For small enough perturbative energy correction,
the corresponding adiabatic time evolution of
the stationary state of the interacting bosons closely follows the one 
of the noninteracting bosons.

Accordingly, under the weak interparticle interaction condition,
the eigenstates of the interacting bosons can be labeled 
by the quantum numbers of the noninteracting bosons.
For example, the ground state of the initial and final points
of the adiabatic cycle $C$ may be denoted as
$\ket{1^{\otimes N}(\lambda)}$, whose overlapping integral with
the unperturbed state $\ket{1^{\otimes N}}$ is large. Also, 
$\ket{1^{\otimes N}(\lambda)}$ 
can be constructed by the
standard perturbation theory with a small parameter $\lambda$.

On the other hand, even a weak interparticle interaction 
can strongly influence the parametric evolution of energy levels 
in the vicinity of level crossings by making avoided crossings.
Hence we need to closely examine the level crossing of the non-interacting 
Bose particles.

In the following, we
argue that the adiabatic time evolution closely follows the one
in the noninteracting system examined above, if the number of the particle
is large enough.
The key is the selection rule for the matrix element 
of ${V}$ in the adiabatic representation in the vicinity 
of the level crossings of non-interacting Bosons.

%

\subsection{``Tunneling'' and direct contributions
  of the interaction in $N=2$}
We show that the effect of the interparticle interaction is 
significantly different, depending on whether a level crossing 
locates either
in $C_{II}$, or in $C_I \cup C_{III}$, 
as for the two body case.
In the former case, the relevant matrix elements may be small since
it involves only tunneling processes through the impenetrable wall.
On the other hand, in processes $C_I$ and $C_{III}$, the matrix
element cannot be negligible. However, it turns out that there happens
to be no corresponding level crossing that affects the population inversion
whose initial state is the ground state.

\begin{figure}
  \centerline{%
    \includegraphics[%
    	height=0.20\textheight
        ]{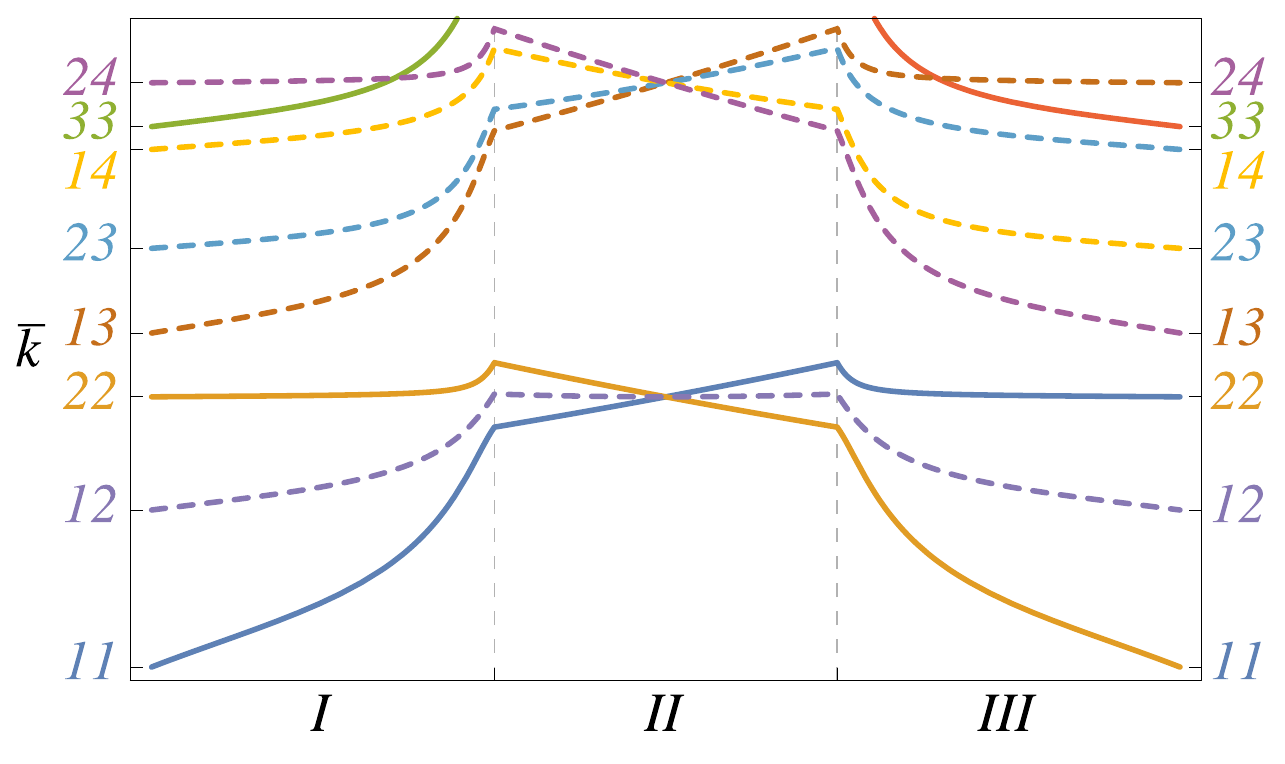}}
  \caption{
    Parametric evolution of normalized wavenumbers for $N=2$.
    Full lines indicate the levels whose initial states are
    $\ket{nn}$ ($n=1,\dots,4$) (see, Figure~\ref{fig:N1}). 
    Other levels are depicted by dashed lines.
    Other parameters are the same as in Figure~\ref{fig:N1}.
    }
  \label{fig:N2}
\end{figure}

The parametric evolutions of eigenenergies of the noninteracting two particle 
system are depicted in Figure~\ref{fig:N2}, in terms of the averaged
wavenumber $\bar{k}$ (see, \eqref{eq:def_bar_k}). The parametric evolution of the eigenenergy that
connects $\ket{11}$ and $\ket{22}$ has a level crossing with two 
eigenenergies during $C_{II}$. The initial states of these energy
levels are $\ket{22}$ and $\ket{12}$, which are
$\ket{L_1L_1}$ and $\ket{R_1L_1}$ during $C_{II}$, respectively.

We examine the matrix elements of the interparticle
interaction term $V$ 
between
the adiabatic basis vectors
$\ket{R_1R_1}$, $\ket{L_1L_1}$ and $\ket{R_1L_1}$. Note that
$\ket{R_1R_1}$ corresponds to the initial state $\ket{11}$
of the adiabatic cycle
\begin{eqnarray}
  \label{eq:V_R1L1R1R1}
  && 
  \bra{R_1,L_1}{V}\ket{R_1,R_1}
  \nonumber \\ &&\quad 
  \!=\! 
  \sqrt{2}\lambda
  \!\!
  \int_0^L 
  \!\!
  \{\psi_{R_1}(x)\psi_{L_1}(x)\}^*\{\psi_{R_1}(x)\}^2dx
  ,
\end{eqnarray}
for example.
Since the single-particle adiabatic eigenfunctions
$\psi_{L_1}(x)$ and $\psi_{R_1}(x)$ are completely localized
in the left and right wells, respectively, the overlapping integral 
is zero, if the $\delta$-wall is completely impenetrable during $C_{II}$.
The level crossing accordingly remains even in the presence of the 
interparticle interaction. Thus the adiabatic cycle $C$ induces
the complete population inversion from $\ket{11}$ to $\ket{22}$,
as in the non-interacting case.

Let us examine the case that the $\delta$-wall during $C_{II}$
allows the tunneling leakage of particles due to some imperfections.
Still, we may expect 
that
the matrix elements due to the tunneling corrections
are exponentially small. Since the resultant energy gap of the avoided
crossing is also exponentially small, we may expect that
the diabatic process easily almost recovers 
the complete population inversion.

Also, during the second process $C_{II}$, the left and right part
of the well may be separated. 
This allows us to make the tunneling correction arbitrarily small.
Accordingly the adiabatic limit 
that follows the extremely small avoided crossing would be difficult
to realize.

On the other hand, if the level crossing appears during $C_I$ 
or $C_{III}$, the interparticle interaction destroys the level
crossing. In Figure~\ref{fig:N2}, such an example is seen 
between the levels whose initial states are $\ket{33}$ and $\ket{24}$,
which is delivered to $\ket{R_2R_2}$ and $\ket{L_1L_2}$, respectively,
in the absence of $V$.

The matrix element
$\bra{33}{V}\ket{24}$ does not vanish in general, since
the relevant single-particle adiabatic eigenfunctions extend
the whole box. Accordingly the level crossings are destroyed
to form avoided crossing. Thus the adiabatic process $C_I$
for example, delivers $\ket{33}$ and $\ket{24}$ at the initial
point of $C_I$, to $\ket{L_1L_2}$ and $\ket{R_2R_2}$, respectively.
This breaks the population inversion whose initial state is a
higher excited state, e.g., the adiabatic cycle $C$ delivers
$\ket{33}$ to $\ket{44}$ in the absence of the interparticle interaction.

\subsection{Selection rule for $N = 3$}
Here, we show that the interparticle interaction do not suppress 
the population inversion for $N > 2$ due to a selection rule of $V$.

\begin{figure}
  \centerline{%
    \includegraphics[%
    	height=0.20\textheight
        ]{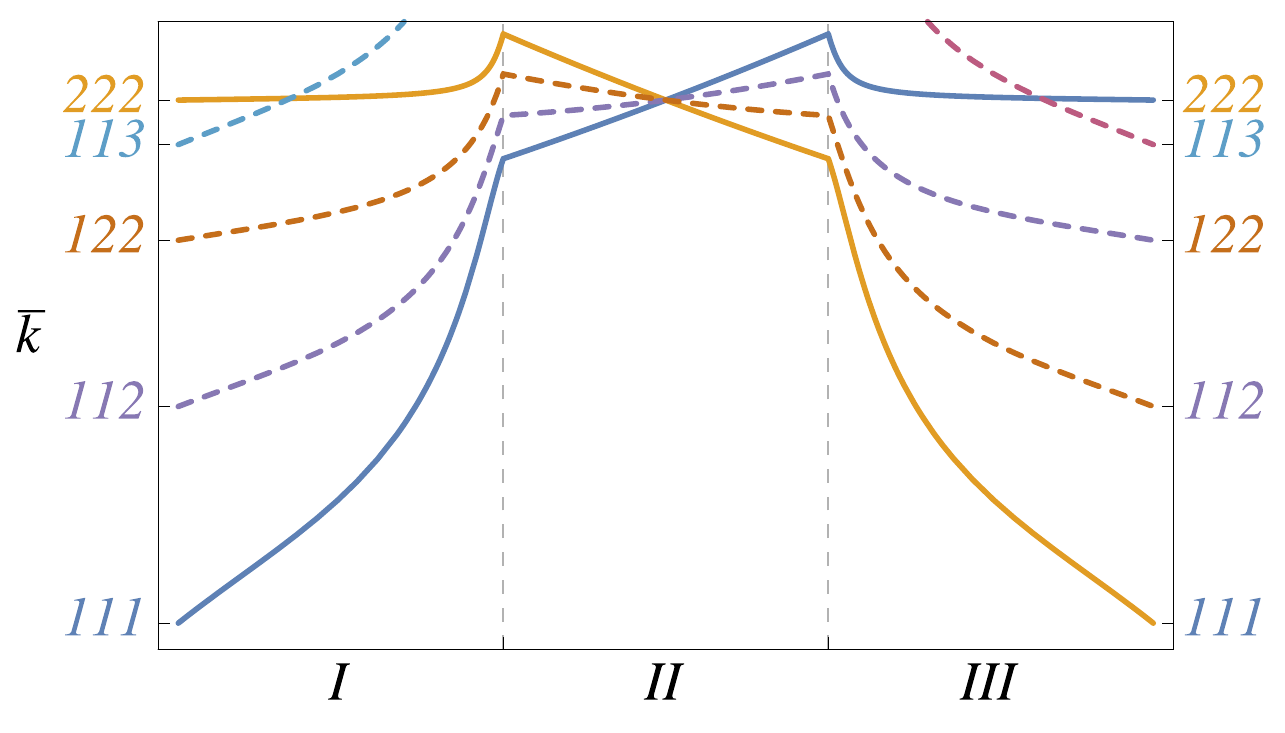}}
  \caption{
    Parametric evolution of normalized wavenumbers for $N=3$.
    Other parameters are the same as in Figure~\ref{fig:N2}.
  }
  \label{fig:N3}
\end{figure}

We explain this with the case $N=3$ (Figure~\ref{fig:N3}).
Let us examine the level whose initial state is 
$\ket{1^{\otimes 3}}$ along $C$. The corresponding final state
is $\ket{2^{\otimes 3}}$ in the absence of the interparticle interaction.

First, the interparticle interaction has no, or exponentially
small effect on the level crossing during $C_{II}$,
as shown in the case of $N=2$.

Second, we examine the level crossing in $C_{III}$, where
the levels whose final states are $\ket{2^{\otimes 3}}$ 
and $\ket{113}$ exhibit crossing. We examine
the matrix element of the interparticle interaction
$\bra{113}{V}\ket{2^{\otimes 3}}$, which vanishes since $V$ is 
a two-body interaction, and the set of quantum numbers
$(1,1,3)$ and $(2,2,2)$ has no common quantum number.


Still, there may be a tiny avoided crossing whose magnitude
can be explained by the standard second-order perturbation theory.
We may expect that the diabatic process
induces the complete population inversion whose final state
is $\ket{2^{\otimes 3}}$. Also, even if the interaction strength $\lambda$
is moderately large, where the topology of the level diagram remains unchanged
except that the avoided crossing becomes noticeable, 
the final state should be $\ket{113}$, whose energy is far larger
than the ground state. In this sense, a incomplete population inversion
should be realized.

\subsection{The population inversion for $N>2$}
\begin{figure}
  \centerline{%
    \includegraphics[%
    	height=0.20\textheight
        ]{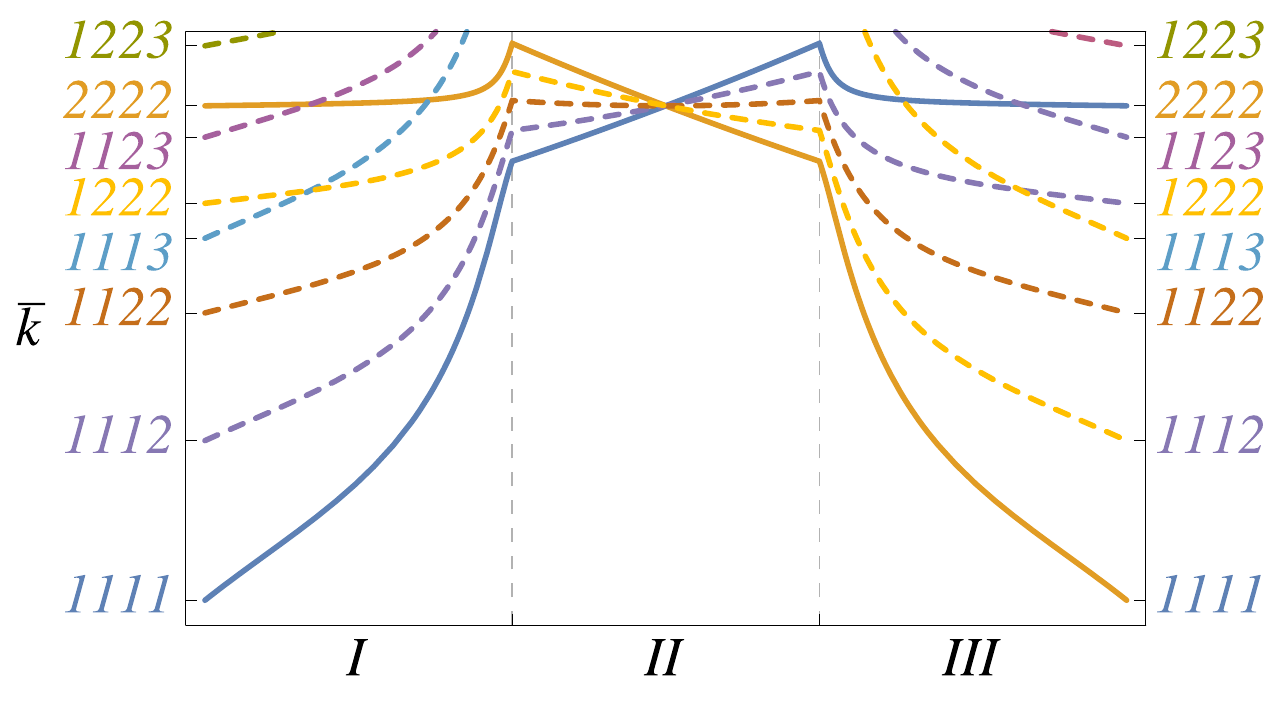}}
  \caption{
    Parametric evolution of normalized wavenumbers for $N=4$.
    Other parameters are the same as in Figure~\ref{fig:N2}.
  }
  \label{fig:N4}
\end{figure}

We 
shall prove 
that the adiabatic cycle $C$ delivers
$\ket{1^{\otimes N}}$ to $\ket{2^{\otimes N}}$ for $N>2$, even 
in the presence of the two-body interparticle interaction.
Here we explain the selection rule for arbitrary $N$ ($>2$), 
and examine each part of the cycle $C$. For example, 
$N=4$ case is shown in Figure~\ref{fig:N4}.

We explain the selection rule of the two-body interparticle interaction 
for $N>2$. Namely, we examine the matrix element
$\bra{n'_1 n'_2 \ldots{}n'_N}{V}\ket{n_1n_2\ldots{}n_N}$.
The matrix element vanishes
when the two sets of quantum numbers
$(n'_1,n'_2,\ldots{},n'_N)$ and $(n_1,n_2,\ldots{}n_N)$
has, at least, three different elements, i.e.,
the number of the common quantum numbers is equal to $N-3$ or less.
In other words, non-vanishing matrix element has
the following 
$\bra{n'_1n'_2n_3\ldots{}n_N}{V}\ket{n_1n_2n_3\ldots{}n_N}$
where $(n_3,\ldots{},n_N)$ are the common quantum numbers.

We examine the first part $C_I$ of $C$. We assume that the system is
initially in the ground state
$\ket{1^{\otimes N}(\lambda)}$.
According to the selection rule, it is sufficient to examine
$\ket{1^{\otimes N-2}\psi,\phi}$, where 
$\ket{\psi}$ and $\ket{\phi}$ are single particle adiabatic states,
e.g. $\ket{2}$. Now we examine whether the eigenenergies of
these states are degenerate. This is equivalent 
to compare the eigenenergies corresponding to 
$\ket{11}$ and $\ket{\psi,\phi}$ of the two particle system. As is seen in
Figure~\ref{fig:N2}, there is no level crossing in $C_I$.
In this sense, there is no effective level crossing
with the level 
$\ket{1^{\otimes N}(\lambda)}$, 
during $C_I$.

As for $C_{III}$, we conclude from a similar argument above, that 
the energy level corresponding to 
$\ket{2^{\otimes N}(\lambda)}$ 
has no effective level crossing.

Next we examine $C_{II}$, where the system is in 
$\ket{R_1^{\otimes N}}$.
According to the selection rule, 
it suffices to examine
$\ket{R_1^{\otimes N-2}\psi,\phi}$ with 
single particle adiabatic states $\ket{\psi}$ and $\ket{\phi}$.
To clarify the level crossing, we compare
$\ket{R_1R_1}$ with $\ket{\psi,\phi}$.
There are three cases.
First, the levels corresponding to $\ket{R_1R_1}$ and 
$\ket{R_n,R_{n'}}$ ($(n,n')\ne(1,1)$) do not occur.
Second, the levels corresponding to $\ket{R_1R_1}$ and 
$\ket{R_n,L_{n'}}$ exhibits a degeneracy only when $n=1$ and $n'=1$,
where the corresponding matrix element involves a single-particle tunneling.
Third, the levels corresponding to $\ket{R_1R_1}$ and 
$\ket{L_n,L_{n'}}$ exhibits a degeneracy only when $n=1$ and $n'=1$,
where the corresponding matrix element involves a two-particle tunneling.
Since the matrix elements involving a tunneling contribution is
exponentially small, the resultant gap should be also small.
Hence the diabatic process should occur even when the speed of the
impenetrable wall is 
moderately
slow.

\section{Discussion and summary}

We here 
argue
that the experimental realization of the population inversion suggested in this paper is feasible with 
the current state of the art.
For example, we may utilize
the scheme~\cite{Meyrath-PRA-71-041604} to realize $\delta$-wall with an approximate Gaussian wall.

Another possibility is to use a heavy particle as a wall, whose position
may be manipulated by, say, an optical tweezer.  
The effective interaction between the wall particle and
other particle may be tuned by external fields.

We note that the present scheme may offer a way to realize 
another exotic nonequilibrium states. Let us suppose, for example, the state of bosons is in 
$\ket{2^{{\otimes}N}(\lambda)}$, which can be generated from the adiabatic cycle $C$.
After the interparticle interaction $\lambda$ is adiabatically increased to $\infty$,
the system arrives the higher excited state of the Tonks-Girardeau system, 
which may be described by the Lieb-Linigher model with the infinite
interparticle interaction 
strength~\cite{Paredes-Nature-499-277,Kinoshita-Science-305-1125}.
Similarly, after $\lambda$ is adiabatically 
decreased to $-\infty$, the system now arrives at the higher excited state of 
the super-Tonks-Girardeau system~\cite{Haller-Science-325-1224}. This 
state is a much
more highly-excited state compared to
the super-Tonks-Girardeau state, 
because
the initial state $\ket{2^{{\otimes}N}(\lambda)}$
is a higher excited state of noninteracting bosons.

In summary, we 
have shown that the adiabatic cycle $C$ induces
the nearly complete population inversion of the 
multi-boson 
system,
when the interparticle interaction is not too strong. As pointed out
in Ref.~\cite{Kasumie-arXiv-151007854} for a single particle case,
the present scheme may be extended to 
the case of an arbitrary shape of the confinement potential $V(x)$.

\ack
This research was supported by the Japan Ministry of Education, Culture, Sports, Science and Technology under the Grant number 15K05216.


%


\end{document}